\documentclass{elsart}%
\usepackage{amsmath}
\usepackage{amssymb}
\usepackage{graphicx,amssymb}
\usepackage[margin=10pt,font=small,labelfont=bf,labelsep=endash]{caption}
\usepackage{amsfonts}
\usepackage{graphicx}%
\providecommand{\U}[1]{\protect\rule{.1in}{.1in}}
\providecommand{\U}[1]{\protect\rule{.1in}{.1in}}
\begin{document}

\title{ }

\begin{frontmatter}
\title{Poisson-Fermi Model of Single Ion Activities}
\author{Jinn-Liang Liu}
\address{Department of Applied Mathematics, National Hsinchu University of Education,
Hsinchu 300, Taiwan. E-mail: jinnliu@mail.nhcue.edu.tw}
\collab{Bob Eisenberg}
\address{Department of Molecular Biophysics and Physiology, Rush University,
Chicago, IL 60612 USA. E-mail: beisenbe@rush.edu}
\begin{abstract}
A Poisson-Fermi model is proposed for calculating activity coefficients of single
ions in strong electrolyte solutions based on the experimental Born radii and
hydration shells of ions in aqueous solutions. The steric effect of water molecules and
interstitial voids in the first and second hydration shells play an important role in
our model. The screening and polarization effects of water are also included in the
model that can thus describe spatial variations of dielectric permittivity, water density,
void volume, and ionic concentration. The activity coefficients obtained by the
Poisson-Fermi model with only one adjustable parameter are shown to agree with experimental data, which vary
nonmonotonically with salt concentrations.
\end{abstract}
\end{frontmatter}

\section{Introduction}

Comprehensive discussions of theoretical and experimental studies on the
activity coefficient of single ions in electrolyte solutions have been
recently given by Fraenkel \cite{F10}, Valik\'{o} and Boda \cite{VB15}, and
Rowland et al. \cite{RK15}, where more references can also be found. The
Poisson-Fermi (PF) model proposed in this paper belongs to the continuum
approach that traces back to the simple, elegant, but very coarse theory ---
the Debye-H\"{u}ckel (DH) theory. As mentioned by Fraenkel, the continuum
theory has evolved in the past century into a series of modified
Poisson-Boltzmann (PB) equations that can involve an overwhelmingly large
number of parameters in order to fit Monte Carlo (MC), molecular dynamics
(MD), or experimental data. Many expressions of those parameters are rather
long and tedious and do not have clear physical meaning \cite{F10}.

The Debye-H\"{u}ckel model is derived from a linearized PB equation
\cite{LM03}. Extended from the DH model, the Pitzer model \cite{P73} is the
most eminent approach to modeling the thermodynamic properties of
multicomponent electrolyte solutions due to its unmatched precision over wide
ranges of temperature and pressure \cite{RK15}. However, the combinatorial
explosion of adjustable parameters in the extended DH modeling functions
(including Pitzer) can cause profound difficulties in fitting experimental
data and independent verification because the parameters are very sensitive to
numerous related thermodynamic properties in multicomponent systems
\cite{RK15}. The Poisson-Fermi model proposed here involves only one
adjustable parameter.

The ineffectiveness of previous Poisson-Boltzmann models is mainly due to
inaccurate treatments of the steric and correlation effects of ions and water
molecules whose nonuniform charges and sizes can have significant impact on
the activities of all particles in an electrolyte system. Unfortunately, the
point charge particles of PB theories have electric fields that are most
approximate where they are largest, near the point. PB theories are not an
appealing choice for the leading terms in a series of approximations, for that
reason. The PF theory developed in our papers \cite{L13,LE13,LE14,LE14b,LE15}
demonstrates how these two effects can be described by a simple steric
potential and a correlation length of ions. The parameters of the PF theory
describe distinct physical properties of the system in a clear way
\cite{LE14b}. The Gibbs-Fermi free energy of the PF model reduces to the
classical Gibbs free energy of the PB model when the steric potential and
correlation length are omitted \cite{LE14b}. The PF model has been verified
with either MC, MD or double layer data at (more or less) equilibrium
\cite{L13,LE13,LE14}, and nonequilibrium data from calcium and gramicidin
channels \cite{LE14b,LE15}.

Here, we apply the PF theory to study the activity properties of individual
ions in strong electrolytes. The steric effect of all particles and the
interstitial voids that accompany them are described by a Fermi-like
distribution that defines the water densities in the hydration shell of a
solvated ion and the particle concentrations in the solvent region outside the
hydration shell. The resulting correlations produce a dielectric function that
shows variations in permittivity around the solvated ion. The experimental
concentration-dependent dielectric constant model proposed in \cite{VB15} is
used to define the concentration-dependent Born radii of the solvated ion in
the present work. The experimental data of the activity coefficients of NaCl
and CaCl$_{2}$ electrolytes reported in \cite{WR04} are used to test the PF model.

\section{Theory}

The activity coefficient $\gamma_{i}$ of an ion of species $i$ in electrolyte
solutions describes the deviation of the chemical potential of the ion from
ideality ($\gamma_{i}=1$). The excess chemical potential is $\mu
_{i}^{\text{ex}}=k_{B}T\ln\gamma_{i}$, where $k_{B}$ is the Boltzmann constant
and $T$ is an absolute temperature. In Poisson-Boltzmann theory, the excess
chemical potential can be calculated by \cite{BC00}%
\begin{equation}
\mu_{i}^{\text{ex}}=\Delta G_{i}^{\text{PB}}-\Delta G_{i}^{0}\text{, \ }\Delta
G_{i}^{\text{PB}}=\frac{1}{2}q_{i}\phi^{\text{PB}}(\mathbf{0})\text{,
\ }\Delta G_{i}^{0}=\frac{1}{2}q_{i}\phi^{0}(\mathbf{0})\text{,} \tag{1}%
\end{equation}
where the center of the hydrated ion (also denoted by $i$) is set to the
origin $\mathbf{0}$ for convenience in the following discussion and $q_{i}$ is
the ionic charge. The potential function $\phi^{\text{PB}}(\mathbf{r})$ of
spatial variable $\mathbf{r}$ is found by solving the Poisson-Boltzmann
equation%
\begin{align}
-\epsilon_{s}\nabla^{2}\phi^{\text{PB}}(\mathbf{r})  &  =\sum_{j=1}^{K}%
q_{j}C_{j}(\mathbf{r})=\rho(\mathbf{r})\text{,}\tag{2}\\
C_{j}(\mathbf{r})  &  =C_{j}^{\text{B}}\exp\left(  -\beta_{j}\phi^{\text{PB}%
}(\mathbf{r})\right)  \text{,} \tag{3}%
\end{align}
where the concentration function $C_{j}(\mathbf{r})$ is described by a
Boltzmann distribution (3) with a constant bulk concentration $C_{j}%
^{\text{B}}$, $\epsilon_{s}=\epsilon_{\text{w}}\epsilon_{0}$, $\epsilon
_{\text{w}}$ is the dielectric constant of bulk water, and $\epsilon_{0}$ is
the vacuum permittivity. The potential $\phi^{0}(\mathbf{r})$ of the ideal
system is obtained by setting $\rho(\mathbf{r})=0$ in (2), i.e., all ions of
$K$ species in the system do not electrostatically interact with each other
since $q_{j}=0$ for all $j$. We consider a large domain $\Omega$ of the system
in which $\phi^{\text{PB}}(\mathbf{r})=0$ on the boundary of the domain
$\partial\Omega$. The ideal potential $\phi^{0}(\mathbf{r})$ is then a
constant, i.e., $\Delta G_{i}^{0}$ is a constant reference chemical potential
independent of $C_{j}^{\text{B}}$.

For an equivalent binary system, the Debye-H\"{u}ckel theory simplifies the
calculation by analytically solving a linearized equation of (2) so that the
potential function $\phi^{\text{PB}}(\mathbf{r})$ becomes a constant
\cite{LM03}%
\begin{equation}
\phi^{\text{DH}}=-\frac{q_{i}\kappa}{4\pi\epsilon_{s}}\text{, \ }\frac
{1}{\kappa}=\left(  \frac{\epsilon_{s}k_{B}T}{\sum_{j=1}^{2}q_{j}^{2}%
C_{j}^{\text{B}}L}\right)  ^{1/2} \tag{4}%
\end{equation}
dependent of the bulk concentration $C_{j}^{\text{B}}$, where $L$ is the
Avogadro constant.

The Poisson-Fermi equation proposed in \cite{LE14b} is%
\begin{equation}
\epsilon_{s}\left(  l_{c}^{2}\nabla^{2}-1\right)  \nabla^{2}\phi^{\text{PF}%
}(\mathbf{r})=\sum_{j=1}^{K+1}q_{j}C_{j}(\mathbf{r})=\rho(\mathbf{r})\text{,
}\forall\mathbf{r}\in\Omega_{s} \tag{5}%
\end{equation}%
\begin{equation}
C_{j}(\mathbf{r})=C_{j}^{\text{B}}\exp\left(  -\beta_{j}\phi^{\text{PF}%
}(\mathbf{r})+S^{\text{trc}}(\mathbf{r})\right)  \text{, \ \ }S^{\text{trc}%
}(\mathbf{r})=\ln\frac{\Gamma(\mathbf{r)}}{\Gamma^{\text{B}}}\text{,} \tag{6}%
\end{equation}
where $S^{\text{trc}}(\mathbf{r})$ is called the steric potential,
$\Gamma(\mathbf{r)}=1-\sum_{j=1}^{K+1}v_{j}C_{j}(\mathbf{r})$ is a void
fraction function, $\Gamma^{\text{B}}=1-\sum_{j=1}^{K+1}v_{j}C_{j}^{\text{B}}$
is a constant void fraction, and $v_{j}$ is the volume of a species $j$
particle (hard sphere). Note that the PF equation includes water as the last
species of particles with the zero charge $q_{K+1}=0$. The polarization of the
water and solution is an output of the theory. The water can be described more
realistically, for example, as a quadrupole in later versions of the theory.
The distribution (6) is of Fermi type since all concentration functions are
bounded above, i.e., $C_{j}(\mathbf{r})<1/v_{j}$ for all particle species with
any arbitrary (or even infinite) potential $\phi(\mathbf{r})$ at any location
$\mathbf{r}$ in the domain $\Omega$ \cite{LE14b}. The Boltzmann distribution
(3) would however diverge if $\phi(\mathbf{r})$ tends to infinity. This is a
major deficiency of PB theory for modeling a system with strong local electric
fields or interactions. The PF equation (5) and the Fermi distribution reduce
to the PB equation (2) and the Boltzmann distribution (3), respectively, when
$l_{c}=S^{\text{trc}}=0$, i.e., when the correlation and steric effects are
not considered.

If the correlation length $l_{c}=2a_{i}\neq0$, the dielectric operator
$\widehat{\epsilon}=\epsilon_{s}(1-l_{c}^{2}\nabla^{2})$ approximates the
permittivity of the bulk solvent and the linear response of correlated ions
\cite{L13,LE13,S06,BS11}, where $a_{i}$ is the radius of the ion. The
dielectric function $\widetilde{\epsilon}(\mathbf{r})=\epsilon_{s}%
/(1+\eta/\rho)$ is a further approximation of $\widehat{\epsilon}$. It is
found by transforming (5) into two second-order PDEs \cite{L13}%
\begin{align}
\epsilon_{s}\left(  l_{c}^{2}\nabla^{2}-1\right)  \Psi(\mathbf{r})  &
=\rho(\mathbf{r})\tag{7}\\
\nabla^{2}\phi^{\text{PF}}(\mathbf{r})  &  =\Psi(\mathbf{r}) \tag{8}%
\end{align}
by introducing a density like variable $\Psi$ that yields a polarization
charge density $\eta=-\epsilon_{s}\Psi-\rho$ of water using Maxwell's first
equation \cite{LE13}. Boundary conditions of the new variable $\Psi$ on the
boundary $\partial\Omega$ were derived from the global charge neutrality
condition \cite{L13}.

To obtain more accurate potentials at the origin $\mathbf{0}$, i.e.,
$\phi^{\text{PF}}(\mathbf{0})$, we need to consider the size and hydration
shell of the hydrated ion $i$. The domain $\Omega$ is partitioned into three
parts such that $\Omega=$ $\overline{\Omega}_{\text{Ion}}\cup\overline{\Omega
}_{\text{Sh}}\cup\Omega_{\text{Solv}}$, where $\Omega_{\text{Ion}}$\ is the
spherical domain occupied by the ion $i$, $\Omega_{\text{Sh}}$ is the
hydration shell of the ion, and $\Omega_{\text{Solv}}$ is the rest of the
solvent domain as shown in Fig. 1. The radii of $\Omega_{\text{Ion}}$ and the
outer boundary of $\Omega_{\text{Sh}}$ are denoted by $R_{i}^{Born}$ and
$R_{i}^{Sh}$, respectively, whose values will be determined by experimental
data. It is natural to choose the Born radius $R_{i}^{Born}$ as the radius of
$\Omega_{\text{Ion}}$ \cite{BC00}. We consider both first and second shells of
the ion \cite{RI13,MP11}. The dielectric constants in $\overline{\Omega
}_{\text{Ion}}$ and $\Omega\backslash\overline{\Omega}_{\text{Ion}}$ are
denoted by $\epsilon_{\text{ion}}$ and $\epsilon_{\text{w}}$, respectively.
\begin{figure}[t]
\centering\includegraphics[scale=0.7]{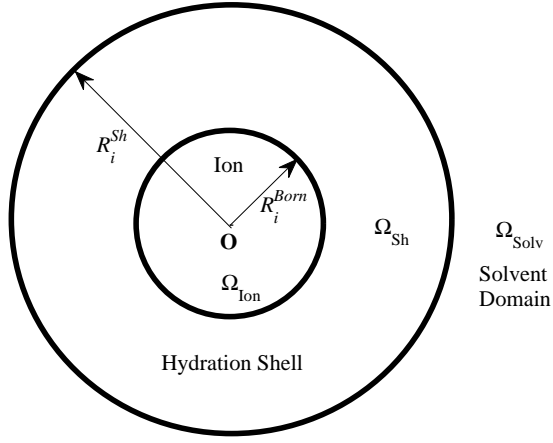}\caption{ The model domain
$\Omega$ is partitioned into the ion domain $\Omega_{\text{Ion}}$ (with radius
$R_{i}^{Born}$), shell domain $\Omega_{\text{Sh}}$ (with radius $R_{i}^{Sh}$),
and solvent domain $\Omega_{\text{Solv}}$.}%
\end{figure}

The PF equation (5) then becomes%
\begin{equation}
\epsilon\left(  l_{c}^{2}\nabla^{2}-1\right)  \nabla^{2}\phi^{\text{PF}%
}(\mathbf{r})=\rho(\mathbf{r})=\left\{
\begin{array}
[c]{l}%
q_{i}\delta(\mathbf{r}-\mathbf{0})\text{ in }\overline{\Omega}_{\text{Ion}}\\
\sum_{j=1}^{K+1}q_{j}C_{j}(\mathbf{r})\text{ in }\Omega\backslash
\overline{\Omega}_{\text{Ion}},
\end{array}
\right.  \tag{9}%
\end{equation}
where $\delta(\mathbf{r}-\mathbf{0})$ is the delta function at the origin,
$l_{c}=0$ in $\overline{\Omega}_{\text{Ion}}$, $l_{c}\neq0$ in $\Omega
\backslash\overline{\Omega}_{\text{Ion}}$, $\epsilon=\epsilon_{\text{ion}%
}\epsilon_{0}$ in $\overline{\Omega}_{\text{Ion}}$, and $\epsilon=\epsilon
_{s}=\epsilon_{\text{w}}\epsilon_{0}$ in $\Omega\backslash\overline{\Omega
}_{\text{Ion}}$. The shell radius $R_{i}^{Sh}$ is determined by Eq. (6) as%
\begin{equation}
S_{\text{Sh}}^{\text{trc}}=\ln\frac{V_{\text{Sh}}-v_{\text{w}}O_{i}^{\text{w}%
}}{V_{\text{Sh}}\Gamma^{\text{B}}}=\ln\frac{O_{i}^{\text{w}}}{V_{\text{Sh}%
}C_{\text{w}}^{\text{B}}}\Rightarrow V_{\text{Sh}}=\frac{\Gamma^{\text{B}}%
}{C_{\text{w}}^{\text{B}}}O_{i}^{\text{w}}+v_{\text{w}}O_{i}^{\text{w}%
}\text{,} \tag{10}%
\end{equation}
where $v_{\text{w}}$ is the volume of a water molecule and $V_{\text{Sh}}$ is
the volume of the hydration shell that depends on the bulk void fraction
$\Gamma^{\text{B}}$, the bulk water density $C_{\text{w}}^{\text{B}}$, and the
total number $O_{i}^{\text{w}}$ (coordination number) of water molecules
occupying the shell of the hydrated ion $i$. Note that the shell volume
$V_{\text{Sh}}$ varies with bulk ionic concentrations $C_{j}^{\text{B}}$. The
occupancy number $O_{i}^{\text{w}}$ is given by experimental data
\cite{RI13,MP11} and so is the shell volume that of course determines the
shell radius $R_{i}^{Sh}$.

To deal with the singular problem of the delta function $\delta(\mathbf{r}%
-\mathbf{0})$ in Eq. (9), we use the numerical methods proposed in \cite{L13}
to calculate $\phi^{\text{PF}}(\mathbf{r})$ as follows:

\begin{description}
\item[(i)] Solve the Laplace equation $\nabla^{2}\phi^{\text{L}}%
(\mathbf{r})=0$ in $\Omega_{\text{Ion}}$ with the boundary condition
$\phi^{\text{L}}(\mathbf{r})=\phi^{\ast}(\mathbf{r})=q_{i}/(4\pi
\epsilon_{\text{ion}}\epsilon_{0}\left\vert \mathbf{r-0}\right\vert )$ on
$\partial\Omega_{\text{Ion}}$.

\item[(ii)] Solve the Poisson-Fermi equation (9) in $\Omega\backslash
\overline{\Omega}_{\text{Ion}}$ with the jump condition $\left[
\epsilon\nabla\phi^{\text{PF}}(\mathbf{r})\cdot\mathbf{n}\right]
=-\epsilon_{\text{ion}}\epsilon_{0}\nabla(\phi^{\ast}(\mathbf{r}%
)+\phi^{\text{L}}(\mathbf{r}))\cdot\mathbf{n}$ on $\partial\Omega_{\text{Ion}%
}$ and the zero boundary condition $\phi^{\text{PF}}(\mathbf{r})=0$ on
$\partial\Omega$, where $\left[  u\right]  $ denotes the jump function across
$\partial\Omega_{\text{Ion}}$ \cite{L13}.
\end{description}

The evaluation of the Green function $\phi^{\ast}(\mathbf{r})$ on
$\partial\Omega_{\text{Ion}}$ always yields finite numbers and thus avoids the
singularity. Note that our model can be applied to electrolyte solutions at
any temperature $T$ having any arbitrary number ($K$) of ionic species with
different size spheres and valences.

\section{Results}

Numerical values of model notations are given in Table 1, where the occupancy
number $O_{i}^{\text{w}}=18$ is taken to be the experimental coordination
number of the calcium ion Ca$^{2+}$ given in \cite{RI13} for all ions $i=$
Na$^{+}$, Ca$^{2+}$, and Cl$^{-}$ since the electric potential produced by the
solvated ion diminishes exponentially in the outer shell region in which a
small variation of $O_{i}^{\text{w}}$ for $i=$ Na$^{+}$ and Cl$^{-}$ does not
affect numerical approximations too much. Obviously the coordination number
may be different for different types of ions and at different concentrations
and so on. We were surprised that we can fit experimental data so well using a
single experimentally determined occupancy number for all ions and conditions.

As discussed in \cite{VB15}, the solvation free energy of an ion $i$ should
vary with salt concentrations and can be expressed by a dielectric constant
$\epsilon(C_{i}^{\text{B}})$ that depends on the bulk concentration of the ion
$C_{i}^{\text{B}}$. Following \cite{VB15}, we assume that
\begin{equation}
\epsilon(C_{i}^{\text{B}})=\epsilon_{\text{w}}-\delta_{i}C_{i}^{\text{B}%
}+\left(  C_{i}^{\text{B}}\right)  ^{3/2} \tag{11}%
\end{equation}
with only one parameter $\delta_{i}$, whose value is given in Table 1, instead
of two in \cite{VB15}. Note that $\epsilon(C_{i}^{\text{B}})$ is a constant
when the dimensionless $C_{i}^{\text{B}}$ is given. It is not a function of a
spatial variable $\mathbf{r}$ like $\widetilde{\epsilon}(\mathbf{r})$. The
parameter $\delta_{i}$ represents the ratio of the factor of $C_{i}^{\text{B}%
}$ to that of $\left(  C_{i}^{\text{B}}\right)  ^{3/2}$ in the original
formula, where the factors of various electrolytes are taken from various
sources of either theoretical or experimental data \cite{VB15}. Our ratios
$\delta_{i}$ in Table 1 are comparable with those given in \cite{VB15}.

The Born formula of the solvation energy can thus be modified as
\begin{equation}
\Delta G_{i}^{\text{Born}}(C_{i}^{\text{B}})=\frac{q_{i}^{2}}{8\pi\epsilon
_{0}\theta(C_{i}^{\text{B}})R_{i}^{0}}\left(  \frac{1}{\epsilon_{\text{w}}%
}-1\right)  \text{, \ \ }\theta(C_{i}^{\text{B}})=\frac{\epsilon
(C_{i}^{\text{B}})\left(  \epsilon_{\text{w}}-1\right)  }{\epsilon_{\text{w}%
}\left(  \epsilon(C_{i}^{\text{B}})-1\right)  }, \tag{12}%
\end{equation}
where $R_{i}^{0}$ is the Born radius when $C_{i}^{\text{B}}=0$ ($\theta(0)=1$)
and $R_{i}^{Born}=\theta(C_{i}^{\text{B}})R_{i}^{0}$ is the
concentration-dependent Born radius used to define $\Omega_{\text{Ion}}$\ in
Fig. 1 when $C_{i}^{\text{B}}\neq0$. The Born radii $R_{i}^{0}$ in Table 1 are
cited from \cite{VB15}, which are computed from the experimental hydration
Helmholtz free energies of these ions given in \cite{F04}. All values in Table
1 are either physical or experimental data except that of $\delta_{i}$, which
is the only adjustable parameter in our model. All these values were kept
fixed throughout calculations.

\begin{center}
$%
\begin{tabular}
[c]{c|c|c|c}%
\multicolumn{4}{c}{Table 1. Values of Model Notations}\\\hline
Symbol & Meaning & \ Value & \ Unit\\\hline
\multicolumn{1}{l|}{$k_{B}$} & \multicolumn{1}{|l|}{Boltzmann constant} &
\multicolumn{1}{|l|}{$1.38\times10^{-23}$} & \multicolumn{1}{|l}{J/K}\\
\multicolumn{1}{l|}{$T$} & \multicolumn{1}{|l|}{temperature} &
\multicolumn{1}{|l|}{$298.15$} & \multicolumn{1}{|l}{K}\\
\multicolumn{1}{l|}{$e$} & \multicolumn{1}{|l|}{proton charge} &
\multicolumn{1}{|l|}{$1.602\times10^{-19}$} & \multicolumn{1}{|l}{C}\\
\multicolumn{1}{l|}{$\epsilon_{0}$} & \multicolumn{1}{|l|}{permittivity of
vacuum} & \multicolumn{1}{|l|}{$8.85\times10^{-14}$} &
\multicolumn{1}{|l}{F/cm}\\
\multicolumn{1}{l|}{$\epsilon_{\text{ion}}$, $\epsilon_{\text{w}}$} &
\multicolumn{1}{|l|}{dielectric constants} & \multicolumn{1}{|l|}{$1$,
$78.45$} & \\
\multicolumn{1}{l|}{$l_{c}=2a_{i}$} & \multicolumn{1}{|l|}{correlation length}
& \multicolumn{1}{|l|}{$i=\text{Na}^{+}$,$\text{Ca}^{2+}$, $\text{Cl}^{-}$} &
\AA \\
\multicolumn{1}{l|}{$a_{\text{Na}^{+}}$, $a_{\text{Ca}^{2+}}$} &
\multicolumn{1}{|l|}{radii} & \multicolumn{1}{|l|}{$0.95$, $0.99$} & \AA \\
\multicolumn{1}{l|}{$a_{\text{Cl}^{-}}$, $a_{\text{H}_{2}\text{O}}$} &
\multicolumn{1}{|l|}{radii} & \multicolumn{1}{|l|}{$1.81$, $1.4$} & \AA \\
\multicolumn{1}{l|}{$R_{\text{Na}^{+}}^{0},$ $R_{\text{Ca}^{2+}}^{0}$,
$R_{\text{Cl}^{-}}^{0}$} & \multicolumn{1}{|l|}{Born radii in Eq. (12)} &
\multicolumn{1}{|l|}{$1.617$, $1.706$, $2.263$} & \AA \\
\multicolumn{1}{l|}{$\delta_{\text{Na}^{+}}$, $\delta_{\text{Ca}^{2+}}$,
$\delta_{\text{Cl}^{-}}$} & \multicolumn{1}{|l|}{in Eq. (11)} &
\multicolumn{1}{|l|}{4.2, 5.1, 3.8} & \\
\multicolumn{1}{l|}{$O_{i}^{\text{w}}$} & \multicolumn{1}{|l|}{in Eq. (10)} &
\multicolumn{1}{|l|}{18} & \multicolumn{1}{|l}{}\\\hline
\end{tabular}
\ $
\end{center}

The PF results of Na$^{+}$, Ca$^{2+}$, and Cl$^{-}$ activity coefficients
agree well with the experimental data \cite{WR04} as shown in Figs. 2 and 3
for NaCl and CaCl$_{2}$ electrolytes, respectively, with various [NaCl] and
[CaCl$_{2}$] from 0 to 2.5 M. In Fig. 4, we observe that the Debye-H\"{u}ckel
theory oversimplifies the Ca$^{2+}$ activity coefficient to a straight line as
frequently mentioned in physical chemistry texts \cite{LM03} because the
theory does not account for the steric and correlation effects of ions and
water, let alone the atomic structure of the ion and its hydration shell as
shown in Fig. 1. Both PB and PF results in Fig. 4 were obtained using the same
atomic Fermi formula (10) for shell radii $R_{i}^{Sh}$ in $\overline{\Omega
}_{\text{Sh}}$ and the same concentration-dependent Born formula (12) for Born
radii $R_{i}^{Born}$ in $\overline{\Omega}_{\text{Ion}}$. Therefore, the only
difference between PB and PF is in $\Omega_{\text{Solv}} $, where
$l_{c}=S^{\text{trc}}=0$ for PB and $l_{c}\neq0$ and $S^{\text{trc}}\neq0$ for
PF. Note that these two formulas are not present in previous PB models. Fig. 4
shows that the correlation and steric effects still play a significant role in
the solvent domain $\Omega_{\text{Solv}}$ although the domain is
$R_{\text{Ca}^{2+}}^{Sh}=4.95$ \AA \ (not shown) away from the center of the
Ca$^{2+}$ ion. The ion and shell domains are the most crucial region to study
ionic activities. For example, Fraenkel's theory is entirely based on this
region --- the so-called smaller-ion shell region \cite{F10}.
\begin{figure}[t]
\centering\includegraphics[scale=0.7]{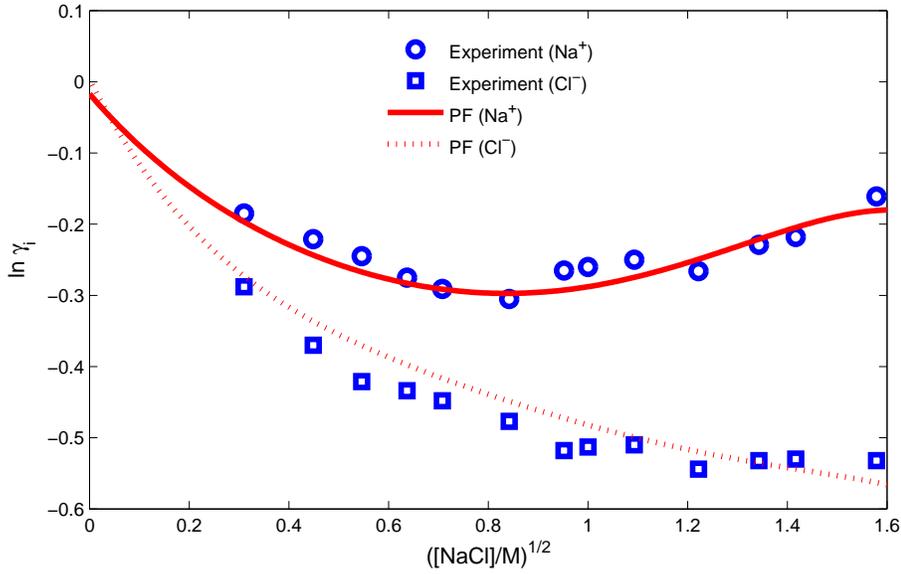}\caption{Comparison of PF
results with experimental data \cite{WR04} on $i=$ Na$^{+}$ and Cl$^{-}$
activity coefficients $\gamma_{i}$ in various [NaCl] from 0 to 2.5 M.}%
\end{figure}\begin{figure}[tt]
\centering\includegraphics[scale=0.7]{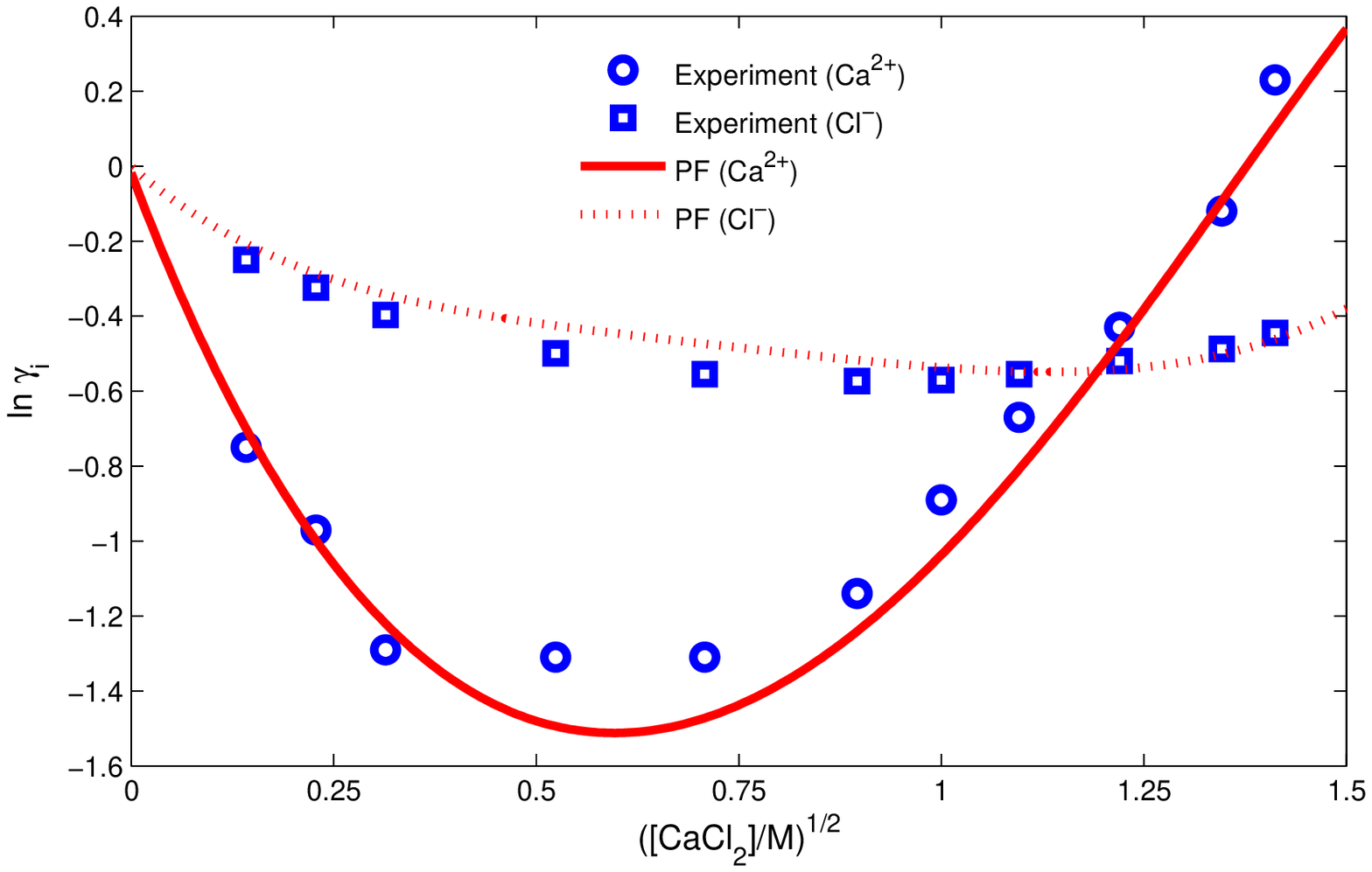}\caption{Comparison of PF
results with experimental data \cite{WR04} on $i=$ Ca$^{2+}$ and Cl$^{-}$
activity coefficients $\gamma_{i}$ in various [CaCl$_{2}$] from 0 to 2 M.}%
\end{figure}\begin{figure}[ttt]
\centering\includegraphics[scale=0.7]{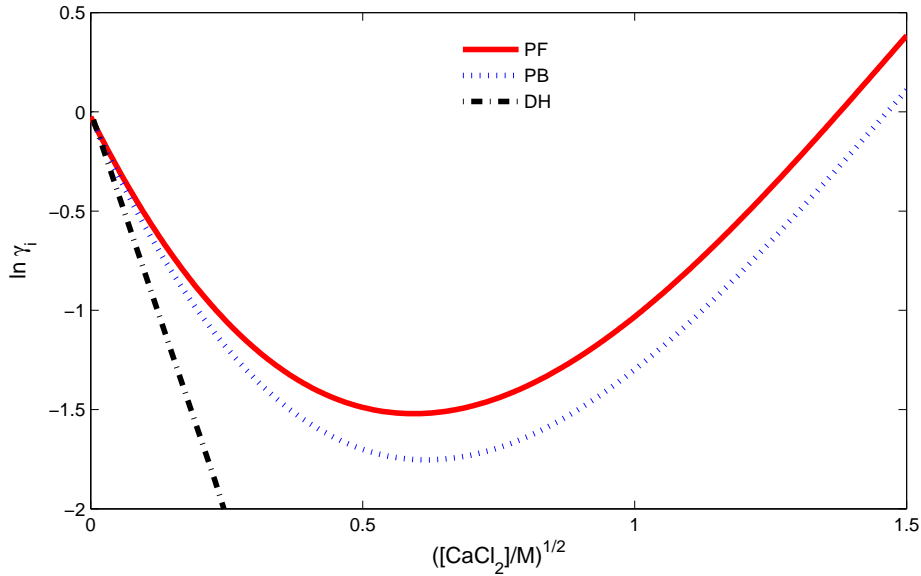}\caption{Comparison of
Poisson-Fermi (PF), Poisson-Boltzmann (PB), and Debye-H\"{u}ckel (DH) results
on $i=$ Ca$^{2+}$ activity coefficients $\gamma_{i}$ in various [CaCl$_{2}$]
from 0 to 2 M.}%
\end{figure}

The PF model can provide more physical details near the solvated ion
(Ca$^{2+}$, for example) in a strong electrolyte ([CaCl$_{2}$] = 2 M) such as
the dielectric function $\widetilde{\epsilon}(\mathbf{r})$ of varying
permittivity (shown in Fig. 5), variable water density $C_{\text{H}_{\text{2}%
}\text{O}}(\mathbf{r})$ (in Fig. 5), concentration of counterion
($C_{\text{Cl}^{-}}(\mathbf{r})$ in Fig. 6), electric potential ($\phi
^{\text{PF}}(\mathbf{r})$ in Fig. 6), and the steric potential ($S^{\text{trc}%
}(\mathbf{r})$ in Fig. 6). Note that the dielectric function $\widetilde
{\epsilon}(\mathbf{r})$ is an output, not an input of the model. The steric
effect is small because the configuration of particles (voids between
particles) does not vary too much from the solvated region to the bulk region.
However, the variation of mean-field water densities $C_{\text{H}_{\text{2}%
}\text{O}}(\mathbf{r})$ has a significant effect on the dielectrics in the
hydration region as shown by the dielectric function $\widetilde{\epsilon
}(\mathbf{r})$. The strong electric potential $\phi^{\text{PF}}(\mathbf{r})$
in the Born cavity $\overline{\Omega}_{\text{Ion}}$ and the water density
$C_{\text{H}_{\text{2}}\text{O}}(\mathbf{r})$ in the hydration shell
$\Omega_{\text{Sh}}$ are the most important factors leading the PF results to
match the experimental data. PF theory deals well with the much more
concentrated solutions in ion channels where void effects are important
\cite{LE14b}. \begin{figure}[t]
\centering\includegraphics[scale=0.7]{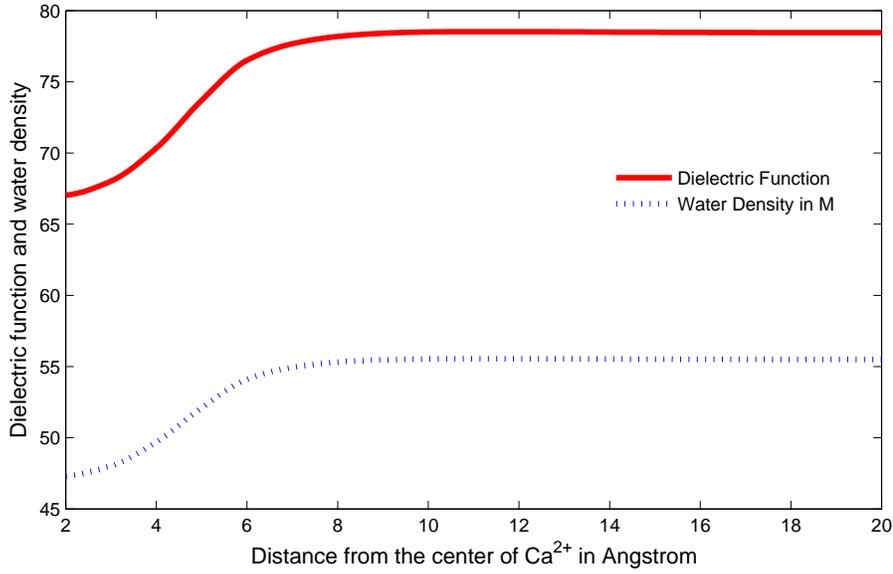}\caption{Dielectric
$\widetilde{\epsilon}(\mathbf{r})$ and water density $C_{\text{H}_{\text{2}%
}\text{O}}(\mathbf{r})$ profiles near the solvated ion Ca$^{2+}$ with
[CaCl$_{2}$] $=2$ M.}%
\end{figure}\begin{figure}[tt]
\centering\includegraphics[scale=0.7]{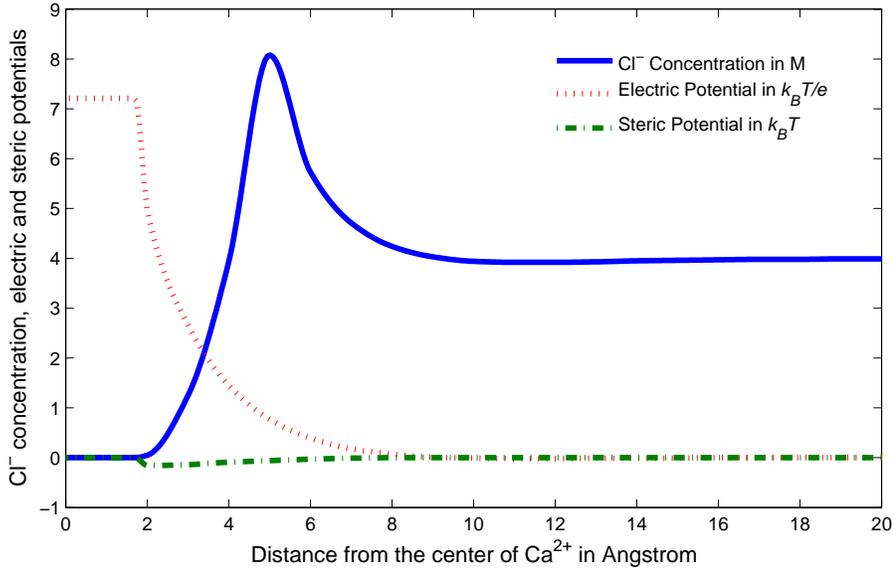}\caption{Cl$^{-}$ concentration
$C_{\text{Cl}^{-}}(\mathbf{r})$, electric potential $\phi^{\text{PF}%
}(\mathbf{r})$, and steric potential $S^{\text{trc}}(\mathbf{r})$ profiles
near the solvated ion Ca$^{2+}$ with [CaCl$_{2}$] $=2$ M.}%
\end{figure}

\section{Conclusion}

We have proposed a Poisson-Fermi model for studying activities of single ions
in strong electrolyte solutions. The atomic structure of ionic cavity and
hydration shells of a solvated ion is modeled by the Born theory and Fermi
distribution using experimental data. The steric effect of ions and water of
nonuniform sizes with interstitial voids and the correlation effect of ions
are also considered in the model. With only one adjustable parameter in the
model, it is shown that the atomic structure and these two effects play a
crucial role to match experimental activity coefficients that vary
nonmonotonically with salt concentrations.

\section{Acknowledgements}

This work was supported in part by the Ministry of Science and Technology of
Taiwan under Grant No. 103-2115-M-134-004-MY2 to J.L.L.

\end{document}